\documentclass{emulateapj}
\usepackage{apjfonts}

\newcommand{\msun}{\ensuremath{M_{\odot}}}			%  Msun
\newcommand{\lsun}{\ensuremath{L_{\odot}}}			%  Lsun
\newcommand{\ks}{K\ensuremath{_{\mathrm{s}}}}		%  Ks
\newcommand{\hh}{H$_{2}$}			%  H2
\newcommand{\water}{H$_{2}$O}		%  H2O
\newcommand{\feii}{\ion{Fe}{2}}		%  FeII
\newcommand{\hii}{\ion{H}{2}}			% HII
\newcommand{\kms}{\textrm{km~s}\ensuremath{^{-1}}}	%  km s-1
\newcommand\rmxaa{\ref@jnl{Rev. Mexicana Astron. Astrofis.}}%
\makeatletter
\def\fnum@figure{{\footnotesize\bf Figure\space\thefigure.~}}
\def\fnum@table{{\bf Table~\thetable}}
\makeatother

\shorttitle{Precessing Jet in Cep~A}
\shortauthors{Cunningham, Moeckel, \& Bally}
\begin{document}

\def\fnum@figure{{\footnotesize\scshape ~~Fig.\space\thefigure.---}}

\title{A PULSED, PRECESSING JET IN CEPHEUS A}
\author{Nathaniel J. Cunningham\altaffilmark{1}, Nickolas Moeckel\altaffilmark{2,3}, John Bally\altaffilmark{2}}
\affil{$^1$Department of Physics and Astronomy, University of Nebraska-Lincoln,116 Brace Laboratory, Lincoln, NE 68588-0111}
\email{ncunningham2@unl.edu}
\affil{$^2$Center for Astrophysics and Space Astronomy,University of Colorado, 389 UCB, Boulder, CO 80309-0389}
\affil{$^3$School of Physics \& Astronomy, University of St Andrews, St Andrews KY16 9SS, Scotland}
\email{Nickolas.Moeckel@colorado.edu}
\email{John.Bally@colorado.edu}

\begin{abstract}
We present near-infrared \hh{}, radio CO, and thermal infrared observations of the nearby massive star-forming region Cepheus A (Cep A).  From \hh{} bow shocks arranged along four distinct jet axes, we infer that the massive protostellar source HW2 drives a pulsed,  precessing jet that has changed its orientation by about 45\arcdeg{} in roughly $10^{4}$ years.  The current HW2 radio jet represents the most recent event in this time series of eruptions.   This scenario is consistent with the recent discovery of a disk around HW2, perpendicular to the current jet orientation, and with the presence of companions at projected distances comparable to the disk radius.  We propose that the Cep~A system formed by the disk-assisted capture of a sibling star by HW2.  We present a numerical model of a 15 \msun{} star with a circumstellar disk, orbited by a companion in an inclined, eccentric orbit.  Close passages of the companion through or near the disk result in periods of enhanced accretion and mass loss, as well as forced precession of the disk and associated orientation changes in the jet.  The observations reveal a second powerful outflow that emerges from radio source HW3c or HW3d.  This flow is associated with blueshifted CO emission and a faint \hh{} bow shock to the east, and with HH~168 to the west.  A collision between the flows from HW2 and HW3c/d may be responsible for X-ray and radio continuum emission in Cep A West.
\end{abstract}


\section{Introduction}

Collimated, bipolar outflows accompany the birth of young stars from the earliest stages of 
star formation to the end of their accretion phase \citep[e.g.][]{reipurth2001,bally2006}.  
The structure  and kinematics of these  flows provide a fossil record of the mass-loss 
histories of the associated young stellar objects (YSOs); the most distant shocks trace the oldest
major mass-loss events while inner jets and shocks provide clues about the recent processes.  The
structure and symmetries of outflows record orientation changes of the underlying accretion
disk  and motion of the outflow sources relative to the surrounding interstellar
medium (ISM).    For example, C-shaped outflows may indicate  deflection by a 
side-wind  (e.g., within expanding plasma of the Orion Nebula, \citealt{bally2006}); 
other C-shaped bends may result from motion of the source YSO through the medium, 
possibly ejected by dynamical processes in a young cluster 
(e.g., HH~498 in NGC 1333, \citealt{bally01}; HH~366 in Barnard 5, \citealt{yu1999}).
Point-symmetric S- or Z-shaped bends can be produced by the precession of a circumstellar disk.
Examples of this kind of symmetry include HH~199 in the L1228 cloud \citep{bally1995} 
and 20126+4104 \citep{su2007}.   Such outflow orientation changes may provide indirect 
evidence for a companion star in  an orbit inclined relative to the disk, inducing forced precession.   

In this paper, we present images of shocked outflows in the Cep A star-forming 
region, together with  radial velocity maps of CO emission.  We interpret 
the outflow morphology as evidence that the massive protostar HW2 drives a 
pulsed, precessing jet  whose orientation changes may be induced
by the periastron passage of a moderate-mass companion in an eccentric, non-coplanar orbit.  
We present a numerical model that demonstrates the plausibility of jet precession forced by a 
companion for the case of HW2.

\subsection{Overview of Cep~A}

The Cep~A star-forming complex contains the second nearest region of massive star formation,
after the Orion complex.   Located at a distance of 725 pc \citep{blaauw1959,crawford1970},  
the Cepheus OB3 association contains a 20 by 60 pc molecular cloud that houses 
six localized peaks of CO emission, designated Cep A through F 
\citep{sargent1977, sargent1979}.  Cep~A  contains dense molecular clumps  
\citep{torrelles1993}, molecular outflows  \citep{rodriguez1980a,narayanan1996,gomez1999}, 
\water{} and OH masers  \citep{cohen1984}, hyper-compact \hii\ regions  
\citep{hughes1984},  variable radio continuum  sources \citep{garay1996}, 
Herbig--Haro objects  \citep{hartigan1986},  bright shock-excited \hh{} emission 
\citep{hartigan2000}, a cluster of far-infrared (FIR) sources
with a luminosity of  $2.5 \times 10^4$ \lsun{}  \citep{koppenaal1979}, and a cluster of 
Class I  and Class II  YSOs \citep{gutermuth2005}.  The bulk of the region's luminosity likely 
arises from radio sources HW2 and HW3c/d \citep{hughes1984} that are associated with bright 
\water{} masers. 

Cep A contains a massive bipolar molecular outflow aligned primarily east--west 
\citep{rodriguez1980}, but with additional components aligned northeast--southwest \citep{bally1990,torrelles1993,narayanan1996,gomez1999} .    The central 2\arcmin{} region contains 
high velocity (HV) as well as more compact extremely high velocity (EHV) CO components 
with radial  velocities ranging from -50 to 70~\kms{} relative to the CO centroid 
\citep{narayanan1996}.  The axis  of the EHV outflow  is rotated roughly 40\arcdeg{} 
clockwise relative to the HV outflow on the plane of the sky. The smaller spatial 
extent together with the higher velocity suggests that the EHV flow traces  a 
younger  outflow component.  Several self-absorption dips follow trends seen in the 
low-velocity  line wings, with  regions east of HW2 blueshifted and west of HW2 redshifted.  
Thus  cooler,  self-absorbing  gas traces the low-velocity bipolar outflow
rather than the quasi-stationary ambient medium.   At low velocities, there are additional 
blue- and redshifted components centered on HW2 that are oriented northeast--southwest.

The Cep~A outflow complex contains several Herbig--Haro objects, including the 
extremely bright HH~168 located about 90\arcsec{} due west of HW2, and several fainter bow 
shocks  located to the east \citep{hartigan2000}.     Fainter HH objects (HH~169 and 174) are
located in the eastern, blueshifted lobe.   Near-inrared (NIR) images show an extremely bright 
reflection nebula centered on HW2 with an illumination cone that opens toward the northeast.
The 2.12 $\mu$m \hh{} line exhibits a complex, filamentary structure.  

There are multiple luminous sources in the complex of HW radio sources
at the core of the Cep~A region, led by HW2, and including HW3c, HW3d, and perhaps 
HW3a, HW8, and HW9. To the west, it has been proposed that radio source W-2 may also 
be internally heated.  Because some of these  (HW2, HW3d, W-2) have the radio 
signatures of jets,  these are likely candidates for driving the outflows observed in CO and
other molecules as well as the shocks traced by \hh{}, \feii{}, and optical
HH emission. The complexity of the outflows, their multiple orientations,
ages, and velocities, and the unclear morphology of some of the shock features make
Cep~A challenging to interpret.  Are deflections involved, as suggested by \citet{goetz1998}?   
Are there additional outflow sources we have not yet detected?  Are
we confusing externally shocked or jet-heated features with self-luminous sources?

HW2 is  the strongest radio continuum  source in the region with a flux density of 15.8
$\pm$ 0.3 mJy at 14.9 GHz \citep{garay1996}; it may be the most luminous source
in Cep~A with $L \approx 10^4$~\lsun{}, implying a mass of 15--20 \msun\@.   
HW2 is located near the center of
a small ($< 5$\arcsec{}), dense clump apparent in the NH$_3$
maps of \citet{torrelles1993} at the tip of one of the larger NH$_3$ concentrations.
Multifrequency observations by \citet{rodriguez1994} using the VLA
show that HW2 is elongated, and that both its spectral index
and size versus frequency relation match those expected for a biconical thermal jet.   Higher
angular resolution observations \citep{hughes1995,hoare1995} show
that HW2 contains a clumpy radio continuum jet whose knots have proper motions
of order 500~\kms{} \citep{curiel2006}.  Many models of the region postulate
that HW2  is the major or sole source of outflows in the region.

The recent interferometric observations of the dust continuum, free--free emission, and
several molecular tracers with arcsecond and subarcsecond angular resolution 
have shown that HW2 is surrounded by a complex dust distribution and several 
close-by protostars.     A hot core that must contain a moderate-mass YSO is located  
about 400 AU  east of  HW2 \citep{martin-pintado2005}.   VLA observations reveal 
a supposedly low-mass protostar that emits radio waves and is located at the center 
of the \water{} maser arc detected by \citet{torrelles2001b}.   HW2 is surrounded 
by a circumstellar disk at least several hundred AU in radius and likely by several 
low-mass protostars \citep{rodriguez2005a,patel2005,brogan2007, comito2007, jimenez-serra2007,torrelles2007}.

The radio sources Cep~A HW3d and HW3b are also associated with a clusters of \water{} 
masers.  HW3c and/or HW3d are suspected to be moderate-mass protostars.  They
are located 5\arcsec{} south of HW2 and close to the western end of a nearly continuous 
chain of radio sources along the southern rim of the Cep~A East radio source complex.   
The eastern part of this chain may in part trace a radio continuum jet from HW3c/d at about 
P.A. $\approx$ 100\arcdeg{}.   The counter-jet direction points directly at the radio source
W-2 at the eastern end of HH~168 in Cep~A West.

\begin{deluxetable}{llccc}
\centering
\tablecaption{NIC-FPS observations of Cepheus A\label{tbl:cepaobs}}
\tablehead{\colhead{Date}		&	\colhead{Filter}		&  \colhead{Number of}&\multicolumn{2}{c}{Exposure time (s)}	\\
\colhead{} & \colhead{} & \colhead{exposures} & \colhead{each} & \colhead{total}}
\startdata
2005 Dec 11	&	\hh-2.12 \micron{}		& 10		&	240	&	2400\\
			&	\ks					& 5		&	20	&	100	\\
2006 Jan 12	&	\hh-2.12 \micron{}		& 5		&	240	&	1200\\
			&	\ks					& 20		&	12	&	240	\\
2007 Jan 29	&	J					& 15		&	20	&	300\\
			&	H					& 15		&	20	&	300\\
			&	\ks					& 15		&	20	&	300\\
\enddata
\end{deluxetable}

Submillimeter wavelength continuum  interferometry at 875 $\mu$m \citep{brogan2007}
shows strong dust emission from HW2, HW3c, and extended emission to the southwest
near HW3a.  No submm continuum was detected from HW3d, leading these authors to
conclude that HW3c is most likely to harbor the second most luminous and massive YSO
in the Cep A core.   In this interpretation, radio source 3d may trace part of a thermal jet
from HW3c.  The associated maser emission may be an indicator of shocks in a dense
molecular gas.   For the rest of this paper, we will assume that the dominant energy
source is embedded in HW3c, but our interpretation does not depend critically on which
peak, HW3c or 3d, contains the source.

The extinction to the HW sources is extremely large with 
$A_{V}\approx$ 500 to 1000 magnitudes.  Thus, none of these radio continuum
sources is visible in the NIR.  An extremely bright infrared (IR) reflection nebula
emerges from the vicinity of HW2 toward P.A. $\approx$ 60\arcdeg{}.
The IR continuum  polarization vectors indicate that the illumination is
coming from the direction of HW2, HW3, and HW8 \citep{colome1995}.    However, 
the accuracy of   polarization measurements is insufficient to determine the source 
unequivocally.

\section{Observations}
\subsection{NIR Imaging}
\begin{figure*}
\plotone{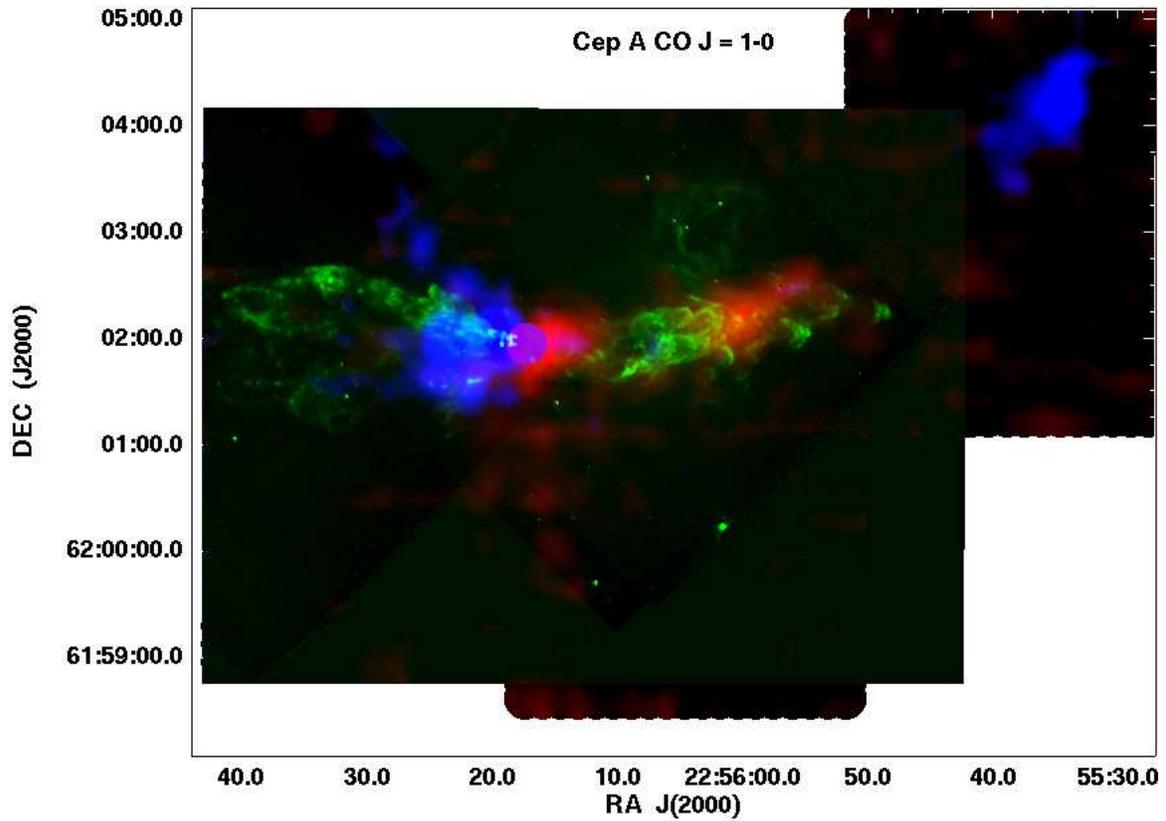}
\caption{The 2.12 \micron{} \hh{} emission (green) superimposed on emission from high-velocity CO with  $-27 < V_\mathrm{LSR} < -18$ \kms{} (blue)
and $-3 < V_\mathrm{LSR} < 6$ \kms{} (red).   In this (and the next
two figures) the purple circle shows a beam-sized spot at the location of HW2.\label{fig:co_hi}
}
\end{figure*}

\begin{figure*}
\plotone{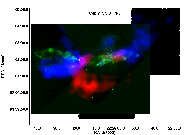}
\caption{The 2.12 \micron{} \hh{} emission (green), and emission from moderate-velocity CO with $-18 < V_\mathrm{LSR} < -13$ \kms{} (blue)
and $-6 < V_\mathrm{LSR} < -3$ \kms{} (red).\label{fig:co_med}
}
\end{figure*}
We observed Cep~A on three nights from 2005 to 2007, using the Near Infrared 
Camera/Fabry Perot Spectrograph (NIC--FPS) on the Astrophysical Research 
Consortium 3.5m telescope at Apache Point Observatory, New Mexico.  The 
NIR emission in the Cep~A region spans 9\arcmin{} east to west, and 
requires at least two telescope pointings for complete coverage with the 
$4.6$\arcmin{}$\times4.6$\arcmin{} field-of-view of NIC--FPS.  On each occasion, the total 
exposure time was evenly split between a western field and an eastern field.  
We obtained broadband images in J, H, and \ks{} filters of the Mauna Kea 
filter set, and in a narrowband (0.4\% bandpass) filter centered on the \hh{} 2.12 \micron{} 
line.  The dates and details of observations taken on each night are listed in 
Table \ref{tbl:cepaobs}.  

Images were taken in dithered sets of 5 or 15 frames; individual frames were reduced 
by performing dark and sky subtraction, flat fielding, and correction for instrumental 
geometric distortion.  Frames in each set were then co-aligned and median combined 
to eliminate bad pixels and to reduce noise.  Registration to absolute coordinates was 
achieved by matching locations of many point sources in each combined image to 
stellar J2000 coordinates for the region from the 2MASS project.  Once properly 
registered to this common frame, all the images taken in a single filter were co-added; 
each region of the combined image was scaled to properly account for a varying 
number of overlapping images and differing exposure times.

\begin{figure*}
\plotone{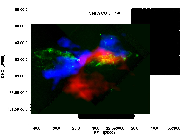}
\caption{The 2.12 \micron{} \hh{} emission (green), and emission from low-velocity CO with $-15 < V_\mathrm{LSR} < -13$ \kms{} (blue)
and $-8 < V_\mathrm{LSR} < -6$ \kms{} (red).\label{fig:co_low}
}
\end{figure*}

As the \hh{} 2.12 \micron{} narrowband filter falls within the bandpass of the \ks{} filter, 
the latter was used as a proxy for the continuum  that contaminates the \hh{} emission 
in the narrowband image.  We removed the continuum component by subtracting a 
scaled version of the final \ks{} image from the same nights as the \hh{} observations.  
The scale factor was chosen based on the best continuum removal as judged by eye 
in the subtracted image.  Variations in seeing between the narrowband and broadband 
exposures result in the incomplete removal of some stars.

\subsection{Optical spectroscopy}

Visual wavelength long-slit spectra of the shocks located east and northeast of Cep~A HW2
were obtained with the Double Imaging Spectrograph (DIS) at the f/10  Nasmyth focus of the
3.5 meter reflector at the Apache Point Observatory on 2007 January 9.    A 1.5\arcsec{} wide 
by 5\arcmin{} long slit was used with a grating providing a spectral resolution 
$R = \lambda / \Delta  \lambda$ $\approx$ 5,000.     At each slit orientation
three images with exposure times of 10 minutes were obtained.

\subsection{Radio observations}

The Nobeyama Radio Observatory (NRO) 45 m radio telescope was used in February 1987
to map the distribution of J=1--0 $^{12}$CO from the Cep~A 
outflow complex with 15\arcsec{} resolution at 115 GHz.    Nobeyama spectra
were collected on a uniform grid with a 10\arcsec{}
spacing over a 5 by 9 arcmin region containing the Cep A outflow complex.  
Over 1900 individual positions were observed for an average integration time of about 1 min
per position, with  20 seconds on source and 20 seconds off source plus 10
seconds of dead time used for reading the spectrometer, and moving the
antenna after each 20 second integration period.  Approximately 14 separate
10-hr periods were devoted to observing Cep A.
A cooled Schottky receiver with a single sideband receiver temperature
$T_{SSB} \approx 350$K was used in conjunction with both
a high (25 kHz/channel) and a low resolution (100 kHz/channel)
2048 channel acousto-optic spectrometer (AOS).  All data
were calibrated using the standard chopper wheel method.  Pointing
was monitored every 2 hr by observations of 43 GHz SiO maser
emission from nearby late-type giant stars using a separate
receiver system.  Both high- and low-resolution AOSs were used to record spectra.  The noise temperature of this
receiver fluctuated between $T_{SSB} \approx 800$K and infinity
due to a problem, so data in these lines are available only over parts of the
source.   Spectra were obtained by position-switching to an
emission-free reference position about 20\arcmin{} to the northwest.
The data were reduced using the COMB package developed by R.~W. Wilson
at AT\&T Bell Laboratories.  

\subsection{Thermal IR Imaging}

Observations of the Cep A region were made using the
Keck Observatory facility mid-IR camera Long Wave Spectrometer (LWS) on 2002 November 16 
November. LWS is a mid-IR imaging and spectroscopy
instrument mounted on the forward Cassegrain focus of Keck I,
employing a Boeing $128 \times 128$ As:Si BIB array with a
$10.2\arcsec{} \times 10.2\arcsec{}$ field of view 
\citep{JonesPuetter1993}.  Weather conditions were poor;
the thermal background and atmospheric transmission varied by  50\%
throughout the first half of the night. Approximately 10 individual
observations (frames) were obtained at slightly overlapping positions 
using the 12.5 \micron\  filter (12 to 13 \micron\  bandpass with $>80$\% transmission). 
The chopping secondary mirror was driven at 2 Hz with a 30\arcsec{}  east--west throw.
Each frame was observed using the standard mid-IR chop-nod
technique with two chopping positions plus (on-source) and
minus (off-source);  after chopping with the source in chop-beam
plus, the telescope was nodded along the chop axis so
that the object would sit in chop-beam minus, and chopping
would continue.   Each frame was observed for one complete
chop-nod cycle, yielding a total on-source integration time of
27.6 s per mosaic frame. The standard stars  $\beta$ Peg,   $\beta$ And, and
$\beta$  Gem were  observed for points-spread
function (PSF) determination and flux calibration.   The images
are nearly diffraction limited at 12.5 $\mu$m with PSF
FWHM = 0.38\arcsec{} and a Strehl ratio of 35\%.  Data reduction
details are given in \citet{shuping2004}.  The image coordinates
are accurate to about 1\arcsec .

\subsection{Observational results and interpretation}

\subsubsection{The CO Outflow Complex}

NRO single-dish observations of accelerated CO emission provide constraints on
the past evolution of the Cep~A outflow complex.  While radio jets and HH objects near their
sources tend to trace the highest velocity and recently ejected outflow components with 
velocities higher than a few hundred \kms{},  HH objects located far from their sources
and shock-excited \hh{} emission tend to trace moderate speeds of 10 to about 200~\kms{}. 
CO and other molecules trace the gas swept up by secondary interactions with the ambient 
medium and have velocities of a few to tens of \kms{}.    Thus, CO and similar tracers tend to
act as calorimeters of the total amount of momentum injected into the parent cloud by the faster
flow components.

The NRO CO data (Figures \ref{fig:co_hi}, \ref{fig:co_med}, and \ref{fig:co_low}) show 
a prominent outflow  complex with complicated structure.   In general, the Cep A CO 
emission is collimated along an east--west direction at the highest velocities, while at 
the lowest velocities, the CO emission is very poorly collimated and bipolar along a 
northeast--southwest direction.  At velocities more than about 8 km~s$^{-1}$ away 
from the rest velocity of the Cep A core ($-11$ \kms{} relative to the local standard of 
rest), the strongest outflow component consists of an east--west bipolar flow that 
is blueshifted toward the east (blue component in Figure \ref{fig:co_hi}) and redshifted 
toward the west.    Additionally, an  anomalous blueshifted lobe appears about 
6\arcmin{} west by northwest of HW2 (upper-right corner of Figure \ref{fig:co_hi}).  
A fainter bipolar component, blueshifted toward the northeast and redshifted
toward the southwest, emerges from the HW2 region at  P.A. $\approx 35\arcdeg$; 
this is most apparent at moderate velocities (diagonal bipolar feature in 
Figure \ref{fig:co_med}). At low velocities of about 2--5 km~s$^{-1}$ with respect 
to the centroid velocity, a prominent blueshifted lobe emerges from near HW3c 
(located about 5\arcsec{} south of HW2) and extends  about 2\arcmin{} east at P.A. 
$\approx 100\arcdeg$.     Below, it is argued that this feature traces an outflow 
from HW3c.

Rather than tracing two independent flows, the brighter east--west outflow from HW2 
and the  secondary, weaker outflow component oriented northeast--southwest at 
P.A. $\approx$ 35\arcdeg{}   may trace  the walls of a large-scale bipolar cavity 
that opens toward the  northeast and southwest.  In this interpretation, the northeast 
cavity walls are at  P.A. $\approx$ 100\arcdeg{} and 40\arcdeg{}  and the southwest 
cavity walls are at P.A. $\approx$ 280\arcdeg{} and 200\arcdeg{}.    The orientation of the
IR reflection nebula emerging from the HW2 region (Figure \ref{fig:cepa_jhk}) provides 
support for this scenario; its axis  lies at P.A. $\approx$ 70\arcdeg{}, which is aligned 
with the opening of the northeast cavity.

\subsubsection{The Core in the Mid-IR}
\begin{figure*}
\plotone{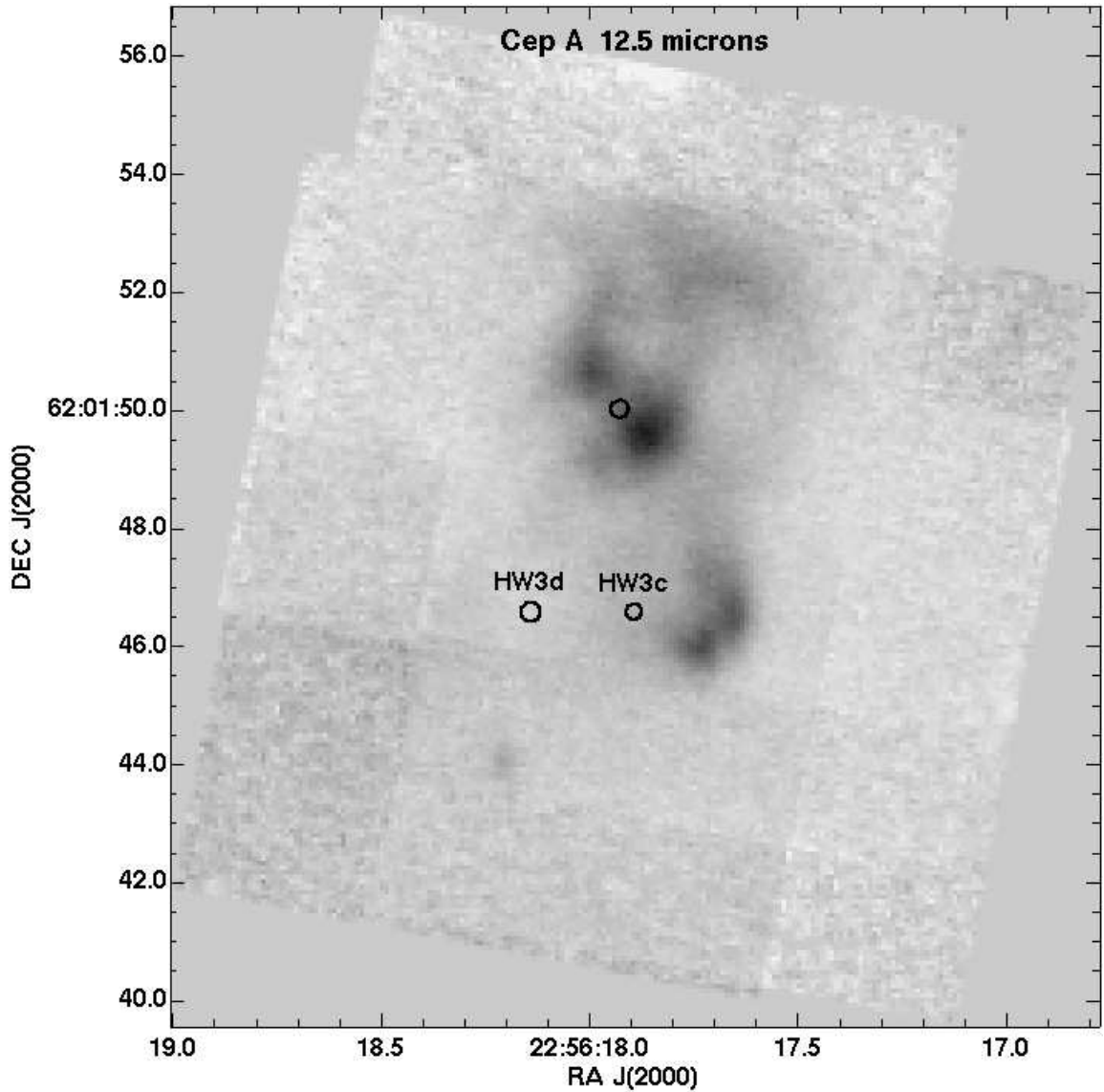}
\caption{The Cep A core at 12.5 $\mu$m observed with LWS on the
Keck 10 m telescope.    The locations of the three suspected most massive 
protostars are shown in black circles.  HW2 is the unlabeled circle.\label{fig:cepa_ir}
}
\end{figure*}

Figure \ref{fig:cepa_ir} shows the Keck 12.5 $\mu$m image of the Cep A core. The 
brightest feature is the double-lobed nebulosity surrounding the position of HW2.  
The two peaks of mid-IR  emission are separated by about 1.4\arcsec{} along an 
axis oriented at P.A. = 45\arcdeg{}, which is similar to the orientation of the HW2 radio 
continuum jet.  The dark lane that separates the two emission peaks has an orientation 
similar to the circumstellar disk surrounding HW2.   However, it is unclear whether the 
dark lane is an IR disk shadow produced by a more compact disk surrounding 
HW2, or  the gap is actually the disk seen in silhouette.  Multicolor imaging is 
needed to determine the nature of the gap.  No prominent mid-IR sources are seen
at the locations of radio sources HW3c and d.  However, an extended IR nebula is 
visible about 2\arcsec{} west of HW3c.  This feature might be illuminated by either 
HW2, 3c, or 3d.  IR polarization measurements are needed to distinguish these possibilities.

A faint point source is located at J(2000) = 22:56:18.2, +62:01:44.  Although there 
are no NIR sources at this location, there is an \water{} maser spot (\water{}-E) 
located 0.7\arcsec{} from the nominal mid-IR source position.

\subsubsection{Radial Velocities of Optical HH Objects}

In contrast to the spectacular 300~\kms{} Doppler shifts seen toward HH~168 in Cep~A West,  
the long-slit spectra of the faint HH objects  HH~169 and 174 located east and
northeast of HW2 show only small Doppler shifts not larger than about 60~\kms{} in 
the core of the line profile.    All components to the east are blueshifted. 
The radio proper motions of the continuum jet emerging from HW2 indicate velocities
of order 500 to 300~\kms{} within a few seconds of the source.  Thus, it may seem 
surprising that shocks located at distances of 1--5\arcmin{}  away show much lower
velocities.  The low velocities either indicate that the flow east of HW2 is mostly along
the plane of the sky, or that the ejecta have been decelerated significantly.   Future
proper motion measurements are needed to distinguish these possibilities.  However, 
most HH objects exhibit rapid decreases in their velocities with 
increasing distance from their sources \citep{reipurth2001, devine1997}.  Such deceleration
is probably caused by the interaction of the ejecta with the ambient medium.    The structure of
the Cep~A cloud is such that the outflows emerging from the Cep~A core propagate relatively
freely toward the west, but impact dense molecular gas toward the east.  Thus, it is
not surprising that HH~168 has high velocities while the eastern shocks associated with 
HH~169 and 174 have low velocities.   For the rest of the paper, we assume that the
space velocities of the ejecta east of the Cep~A core have a speed of about 100~\kms{},
typical of HH objects located at comparable distance from their sources. 

\subsubsection{The \hh{} Outflow Complex}
\begin{figure*}
\plotone{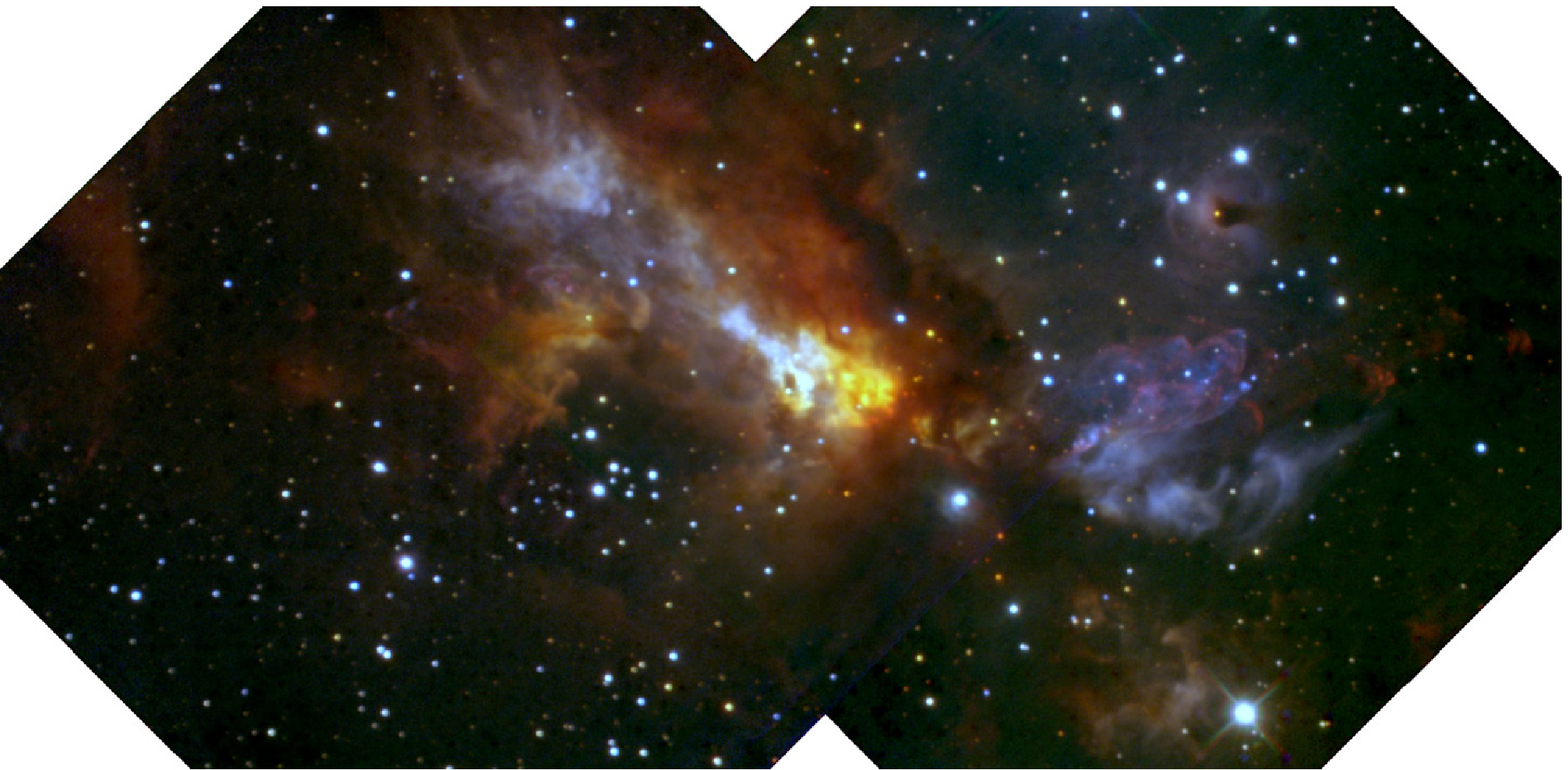}
\caption{Color-combined broadband image of Cep~A\@.  J-band emission is shown in blue, 
H in green, and \ks{} in red.\label{fig:cepa_jhk}}
\end{figure*}

\begin{figure*}
\plotone{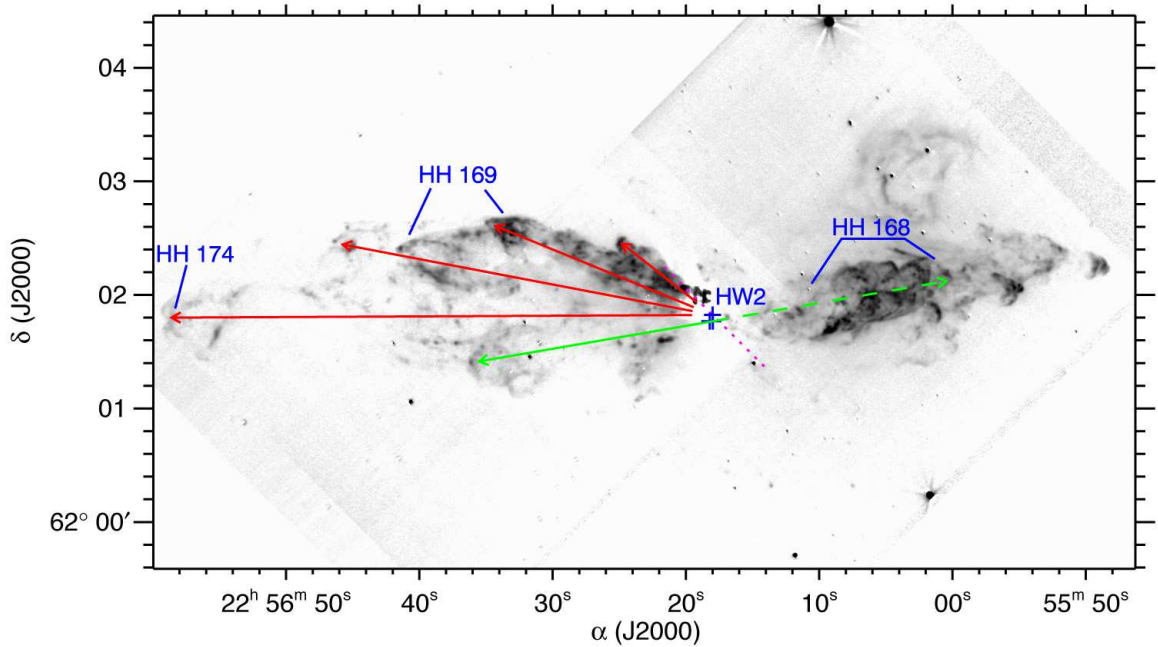}
\caption{Continuum-subtracted 2.12 $\mu$m \hh{} emission in the Cep~A outflow complex.
The successive orientations of a suspected precessing jet from HW2
are indicated by red arrows; the position of HW2 is marked with a cross.  The oldest (eastward) ejection appears to power
HH~174 to the east.   The next two ejections may be
responsible for HH~169. 
The current orientation of the radio jet emerging from HW2 is at position angle (P.A.) = 45\arcdeg{}, indicated by the dashed magenta line.
Bright HH~168, and the \hh{} bows at far right, may result from an outflow from HW3c or HW3d (also marked with crosses) along the green axis.  A faint \hh{} bow marks the opposite lobe of this flow.\label{fig:cepa_h2}}
\end{figure*}

Figure \ref{fig:cepa_jhk} shows a three-color composite image built from the J, H, and \ks{} broadband filter images.    The continuum-subtracted narrowband image of the \hh{} emission in the Cep A complex is displayed in Figure \ref{fig:cepa_h2}, and is shown in relation to the CO emission in Figures \ref{fig:co_hi}--\ref{fig:co_low}.
The narrowband image exhibits three main components:  shock emission east of the central 
protostellar cores containing HW2 and HW3c/d,  bright arcs  of shock-excited emission located
to the west of these cores that is associated with HH~168, and a faint rosette of emission to  the northwest.  The rosette is associated with an east-facing globule or small pillar, located about 1.5\arcmin{} north of HH~168, which exhibits  slightly blueshifted CO emission in Figures \ref{fig:co_med} and \ref{fig:co_low}.  The \hh{} emission is likely to be excited by UV radiation 
from the Cep OB3 association and is  thus probably fluorescent in nature.
The globule contains an $m_\mathrm{B} = 16$ magnitude star that  is also called Cep A IRS2.    
Because there is no redshifted emission in this region, the blueshifted CO emission is unlikely 
to be a tracer of an outflow.  The CO velocity field in the globule may reflect random motions within
the Cep A cloud, or be the result of UV-photo-heating-induced ablation of the globule surface that 
can accelerate gas to a velocity of a few km~s$^{-1}$.

\subsubsection{Bipolar Outflow from HW3c: HH~168 and its Counterflow}

The series of bright arcs and \hh{} bow shocks located 1--3\arcmin{} west of HW2 are 
associated with HH~168. The axes of symmetry of most of these shocks indicate 
a point of origin about 10\arcsec{} south of HW2.   The submillimeter and radio continuum
source HW3c and the bright \water{} maser and radio source HW3d  lie on or near this axis.  
A dim, 40\arcsec{} diameter bow shock
is located on this axis about 2\arcmin{} east of HW3c, directly opposite HH~168.   
This feature and the visual-wavelength  part of HH~168 are symmetrically placed about HW3c.
Thus, HW3c is likely to be the source of these shocks and the associated CO 
outflow components. High-resolution cm-wavelength VLA observations \citep{garay1996}
reveal a chain of radio sources approximately aligned with the axis of this flow; these may either trace an ionized jet, or the interface between a jet and surrounding 
dense cloud material.   

Figures 2 and 3  show that the bipolar outflow from HW3c is associated with low-velocity 
to intermediate-velocity red- and blueshifted gas.      
The bright jet-like \hh{} feature located 30\arcsec\
east of HW3c is associated with the strongest blueshifted CO emission.   However,
about 1\arcmin{} east, the blueshifted CO lobe shifts about 15\arcsec{} south of the 
flow axis and terminates just below (south) of the \hh{} bow shock.  Evidently, 
the HW3c flow interacts strongly with ambient cloud material along its south wall.
However, there appears to be less interaction along the north rim of this cavity, perhaps
because ambient material in this region has been cleared away by the HW2 flow.

The western lobe of this flow is associated with redshifted CO  co-located with HH~168.
Three bow shocks bright in \hh{} lie on the southern side 
of HH~168; their orientations are consistent with a flow powered by HW3c.
The bow shock morphologies and low CO velocities indicate that this flow lies close to
the plane of the sky.    It is somewhat unusual that at visual and NIR wavelengths, 
the redshifted outflow lobe is brighter than the blueshifted lobe.  This feature may be a 
consequence of the morphology of the  Cep A cloud core that appears to be mostly 
located east of the region containing HW2 and HW3c.  The large range of radial velocities 
observed in HH~168 in visual-wavelength spectra may indicate sideways splashing of 
the high velocity ejecta moving mostly in the plane of the sky that has encountered 
slower-moving material. 

\subsubsection{Non-detection of Source W-2 in HH~168 in the IR}

Radio source W  is located at the eastern end of the HH~168 shock complex in Cep~A West.   
The radio emission from this object is extended with an elongation similar to that of the shocked \hh{} 
emission.   Based on radio spectral indices, \citet{garay1996} infer  that subcomponent W-2 houses an 
ultracompact internal source, shielded from view in the IR by the surrounding molecular cloud, 
and that the elongated radio emission is likely due to a jet launched by this source.  Our images
exhibit co-located,  intense NIR \hh{} emission that coincides with bright  optical HH features at the
eastern end of HH~168, making it unlikely that this region is highly obscured.   Yet the broadband
NIR images do not reveal any point sources in this region; thus it seems unlikely that 
a stellar source exists at this location.  Rather,  radio source W may trace a hard shock, with 
speeds above 400~\kms{}.  The detection of soft X-ray emission from this  region \citep{pravdo2005} 
provides support for this interpretation.  Source W is located where two 
outflows, one emerging from HW2 and another from HW3c,  intersect on the plane of the sky.  
Thus it is possible that these flows are actually colliding.  We  suggest  that the radio and X-ray 
emission from HW  may arise  from colliding flows emerging from the two most massive protostars 
in Cep~A East.  This interpretation is generally in line with the westward proper motions measured 
for this radio source \citep{rodriguez2005a}.

\subsubsection{A Pulsed, Precessing Outflow from HW2}

The blueshifted, eastern lobe of the Cep~A outflow contains four distinct chains of 
\hh{} emitting features, each of which terminates in a well-formed bow shock.  The four
chains appear to emerge from the immediate vicinity of HW2.  Their axes, defined by
lines connecting HW2 to the bow shocks at the eastern and north-eastern
ends of the chains,  shift systematically clockwise from nearly  east--west  to northeast--southwest.  
The longest chain, which terminates 4.8\arcmin{} East of HW2 at HH~174, has a 
position angle P.A. $\approx$ 90\arcdeg{}.   The second chain, 
which terminates about 3.4\arcmin{} from HW2 at the eastern component of HH~169, 
has P.A. $\approx$ 80\arcdeg{}.  The third chain terminates at the western component of HH~169,
about 2.2\arcmin{} northeast of HW2 and  has  P.A. $\approx$ 65\arcdeg{} .  The fourth chain ends in a bright
but compact \hh{} bow at P.A. $\approx$ 55\arcdeg{} about 1\arcmin{} from HW2.   The current 
orientation of the HW2 radio jet continues this clockwise migration of outflow orientations
and has  P.A. $\approx$ 45\arcdeg{}.     The chains of \hh{} knots get progressively shorter as the axes
rotate clockwise toward  decreasing P.A.. This remarkable progression may be an indication that HW2
powers a pulsed and precessing jet.   Between each major outflow ejection event, the jet orientation
changes by about 10\arcdeg{}--15\arcdeg{} as seen in projection on the plane of the sky.   

A rough dynamical age for each chain can be estimated by  dividing the length of each 
by the velocity of its tip.   The radial velocities of the HH objects located at the
ends of the first three chains are low ($v < 60$~\kms{}), indicating that 
they are most likely moving close to the plane of the sky.    Fabry--Perot imaging of the \hh{} 
emission \citep{hiriart2004} also indicates low, but chaotic, radial velocities.  
On the other hand, the excitation of the visual wavelength H$\alpha$ and \ion{S}{2} emission of the eastern
HH objects   and comparison with the speeds of other HH objects located at similar
distances from their sources, indicate that shock speeds of order 100~\kms{} are
not unreasonable.   Future proper motion measurements are needed to determine the
velocities  of the ejecta.  Assuming 100~\kms{} \citep[reasonable for similar shocks exhibiting 
NIR \hh{} emission, e.g.][]{reipurth2001}, the dynamical ages of the four eastern \hh{}
chains are 9900, 7000, 4500, and 2100 years, respectively, indicating that an eruption/ejection event  occurs approximately every 2500 years.  Furthermore, the presence of the HW2
radio jet indicates that there is currently an eruption underway.

This periodic, geometric progression suggests an underlying temporal sequence.  We propose 
a model  for Cep~A East in which the accretion disk surrounding HW2,   responsible for 
launching a bipolar  jet along the disk rotation axis, has periodically changed its orientation.  
In this picture,  the easternmost lobe (position angle 90\arcdeg{}) constitutes the oldest 
tracer of past jet activity.   Since its launch, the accretion disk and associated bipolar jet 
at HW2 have been periodically torqued through a series of new orientations.

Periastron  passages of a companion in an eccentric  orbit that is not coplanar with
the disk can readily explain the torques required to explain both the impulsive
nature of the HW2 outflow, and the periodic changes in disk orientation.  A cluster member 
in an elliptical orbit around HW2, with a periastron distance of order the disk radius,  will 
exchange angular momentum with the disk.  If the binary and disk axes are not aligned, 
the result is a progressive tilting of the disk at each periastron passage.  The close passage of 
the orbiting companion will also disturb the disk during periastron passage, triggering an 
increase in the accretion rate onto the central star, and consequent mass loss in the form of
a collimated jet.    The periodic reorientation of the jet axis and disk, and the episodic
character of the outflow, may be the signatures of  a  crowded and dynamically active environment 
surrounding HW2.

The angular extent of the longest shock lobe (position angle 90\arcdeg{}) is 4.8 arcmin, yielding 
a projected length of  about 1 pc given the adopted distance of 725 pc \citep{blaauw1959}.  
As estimated above, the time between disk reorientations (and in this model, the orbital 
period) is $\approx$ 2500 years.   The mass of HW2 is about 15 \msun{} \citep[and references 
therein]{patel2005}, so a less massive companion must have an orbital semimajor axis 
of $\approx$400 AU.  The periastron distance must be somewhat smaller than this value.  
As indicated in the Introduction, Cep~A HW2 has at least two companions
within the required distance, including the ``hot core" that may be heated by a moderate-mass
protostar \citep{martin-pintado2005} and the source of the expanding maser 
ring \citep{torrelles2001b,curiel2002}.

\section{Numerical modeling}

The interaction of a binary companion and a circumstellar disk can
cause the disk to precess \citep[e.g.][]{papaloizou95,terquem99}.
This precession may be traced by the orientation of the jet, and
has been implicated in the appearance of several precessing outflows
\citep[e.g.][]{eisloffel96,davis97}.
During repeated encounters of a captured companion on an eccentric
orbit, the disk orientation moves through several angles impulsively
\citep{moeckel2006}, rather than smoothly varying as would occur in a
nearly circular binary.  Inspired by the observations presented above,
we searched the parameter space of encounters simulated in
\citet{moeckel2007a} and selected three combinations of parameters for
further investigation.

\subsection{Method and initial conditions}

We used a modified version of the SPH/$N$-body code GADGET-2
\citep{springel05} to model interactions between a massive protostar
and a captured companion.  The models and the method are
similar to those of \citet{moeckel2006}, and we summarize them here.
A Keplerian disk is set up with surface density $\Sigma(r) \propto
r^{-1}$, and temperature $T(r) \propto r^{-1/2}$.  This disk is
allowed to evolve in isolation until the system has relaxed from its
initial conditions.  At this time an impactor star is introduced on a
slightly hyperbolic orbit $\sim 3$ disk radii from the primary.  The
stars are modeled as point masses interacting with the gas only
through gravity, and are assigned an accretion radius.  Gas that falls
into the accretion radius and is energetically bound to the star is
removed from the calculation, and its momentum and mass are added to the star.
 The encounter is integrated until the orbital parameters of the two
stars have stopped evolving.

If the orbital timescale of the newly formed binary is much larger
than the orbital period of the disk, the system is analytically
advanced to the next interaction.  Each gas particle is either assigned to
the star that it is most bound to or deemed ejected from the system, giving
us three populations.  The ejected material is removed from the
calculation.  In order to determine a Keplerian orbit, the gas that
remains and the star that it is bound to are treated as a point mass
at their center of mass, and these point masses are analytically
advanced to the next encounter.  Upon restarting the hydrodynamic
simulation, the gas particles are circularized about their star using
a technique described in greater detail in \citet{moeckel2006}.  The
main point of this process is that we avoid a computationally
expensive integration to apastron and back.

\citet{moeckel2007a} performed simulations of encounters
between a 20 \msun{} star with a 2 \msun{}, 500 AU disk and impactors
of varying masses, periastron, and inclination angle.  By inspecting these 
simulations for disk orientation changes of roughly 10\arcdeg{}, we 
selected three impactor parameters to study the HW2 system, 
shown in Table \ref{simulations}.  We scaled the primary and disk 
masses down to values perhaps more suitable for HW2, a 15 \msun{}
primary with 350 AU, 1.5 \msun{} disk.  The mass of the impactor in
all cases is 5 \msun{}; a fairly massive companion is needed to torque
the disk through $\sim 10$\arcdeg{} during a periastron passage.  We
modeled the disk using $\sim 1.28 \times 10^{5}$ particles, and
followed each system through three encounters.

\begin{deluxetable}{cccccc}
\centering
\tablecaption{Simulation parameters.\label{simulations}}
\tablehead{\colhead{$M_\mathrm{primary}$} & \colhead{$M_\mathrm{disk}$} &
\colhead{$r_\mathrm{disk}$} & \colhead{$M_\mathrm{impactor}$} & \colhead{$i$} &
\colhead{$r_\mathrm{peri}$}  \\
           \colhead{\msun}       & \colhead{\msun}      &
\colhead{AU}       & \colhead{\msun}      & \colhead{degrees}&
\colhead{AU}}
\startdata
15  &  1.5  &  350  &  5  &  30  &  150  \\
15  &  1.5  &  350  &  5  &  45  &  80  \\
15  &  1.5  &  350  &  5  &  135  &  80
\enddata
\end{deluxetable}

\subsection{Numerical results}

\begin{figure*}
\plotone{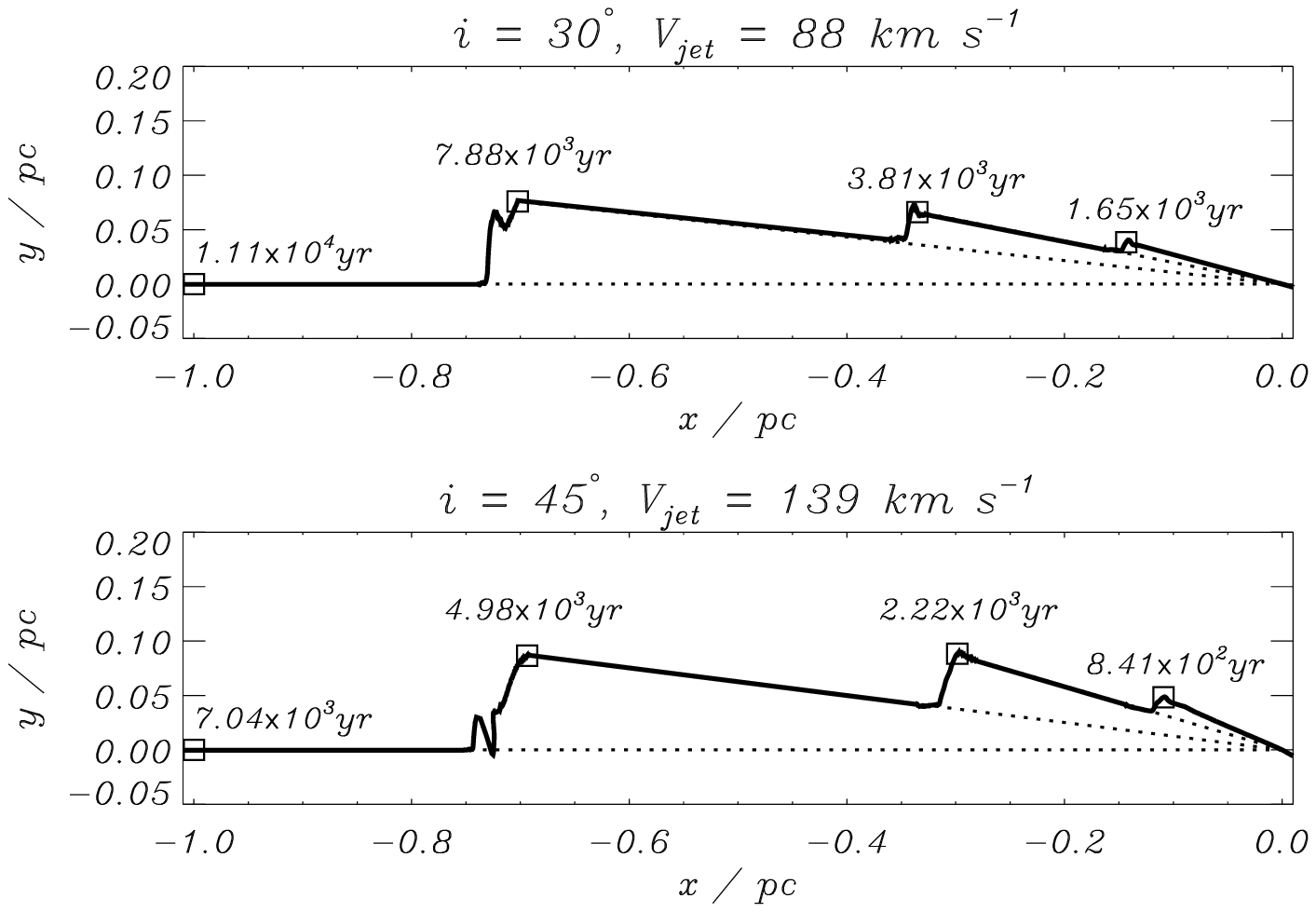}
\caption{Locations of jet ejecta from two simulations, each as viewed at a single time.  The solid lines mark projected distance from outflow source, with older material at far left and recent ejecta at far right.
Dotted lines indicate successive jet orientations; in the absence of disk precession, the jet would remain horizontal.  The jet velocities
have been scaled to give a 1 pc distance to the oldest shock, and the initial jet orientation has been chosen for comparison to Figure
\ref{fig:cepa_h2}.    Boxes indicate where shocks at the ends of the outflow lobes would be
expected, and are labeled with the age of the flow at that point.
\label{fig:jetfigure}}
\end{figure*}

We consider the orientation of the inner disk $\hat{L}_\mathrm{disk}$ to be
defined by the summed angular momenta of the gas particles within 30
AU of the primary.  In order to determine what the outflow appearance
would be, we construct a crude jet model as follows.  We assume that the
orientation of the outflow is the same as the inner disk orientation,
$\hat{V}_\mathrm{jet} \equiv \hat{L}_\mathrm{disk}$.
By assuming a constant-velocity jet launched in this direction, we can
construct a  model of what the jet from this interacting system
will look like.    Of the three cases considered,  the
two prograde simulations ($i = 30$\arcdeg{} and $i = 45$\arcdeg{})
produced jets that are similar in appearance to the observed system.
The retrograde encounter, $i = 135$\arcdeg{}, showed a shift in
orientation similar to the other two for the first passage, but the
subsequent passages did not resemble the actual jet.  We believe this
is in part due to the rapid disk destruction associated with
retrograde passages having small $r_\mathrm{peri} / r_\mathrm{disk}$ 
\citep{moeckel2006}.

Plots of the jets through the first three passages from the two
prograde simulations are shown in Figure \ref{fig:jetfigure}.  They
have been rotated about the initial jet axis in order to most closely match the
observed appearance of the HW2 system, and the initial jet axis lies
on the x-axis to match the orientation of the jet lobes in Figure
\ref{fig:cepa_h2}.  The simulations yield the elapsed time of the
encounters, and at an assumed distance of 725 pc, the distance to the
oldest bowshock is approximately 1 pc.  By choosing the velocity of
the jet to match this distance, we find jet velocities of 88~\kms{}
for the 30\arcdeg{} case, and 139~\kms{} for the 45\arcdeg{} case.
The boxes on the figure indicate where bowshocks would be expected,
and are labeled with the age of the flow at that point.  While both
simulations capture the qualitative orientation of the jet, the case
with $i = 45$\arcdeg{} appears to match the observed jet orientation
slightly better.

We stress that we are not claiming that the parameters studied here
reflect the reality of the HW2 region's dynamics.  The uncertainties in the
masses, radii, and even number of objects impede attempts to
model the actual system, and the range of parameters that
produce similar outflows is large.  Rather, these simulations should
be viewed as a proof-of-concept experiment.  Using plausible values
for the masses of the components, we can generate outflow orientations
similar to those observed, via a series of postcapture interactions
between a massive protostellar system and a moderate-mass companion.

\section{Discussion}

The two most massive protostellar objects in Cep A, HW2 and HW3c,
each power separate outflows.    HW2 appears to have been responsible
for at least four quasi-periodic ejections that produced collimated flows
visible in NIR \hh{} emission in the eastern lobe of the Cep A
outflow complex.     While the first eruption launched
a flow nearly due east, subsequent ejections indicate that the source orientation
rotated north by about 10\arcdeg{} between events.  The current orientation of
the radio jet, rotated clockwise by 45\arcdeg{} compared to the ejection axis of the
first event, continues this trend.      Radial velocity measurements of the 
visual wavelength emission associated with HH 169 and 174 located at
the tips of three of the outflow axes indicate low radial velocities and dispersions,
an indication that the outflow axis probably lies close to the plane of the sky.

The bright NIR reflection nebula 
emerging from the vicinity of HW2 has an axis close to the last two ejection
events.  Millimeter wavelength CO emission toward the eastern lobe of
the Cep A outflow complex reveals two ridges of blueshifted emission
at P.A. $\sim$ 100\arcdeg{} and 35\arcdeg{}, approximately symmetrically placed
about the axis of the reflection nebula.   A pair of redshifted ridges of
CO emission are located on the opposite side of HW2 toward the southwest.
These CO features may define the walls of an outflow cavity created by
the outflow over time.    The CO outflow from HW3c is superimposed on
the east--west rim of this suspected cavity.

The data presented above are interpreted in terms of a pulsed, 
precessing jet  emerging from Cep A HW2.   
A numerical model is used to show that a pulsed, precessing jet can be the result
of disk perturbations produced by a companion star in an eccentric orbit that is
not co-planar with the disk.   This scenario is the most likely outcome of the
capture-formed binary proposed by \citep{moeckel2007a,moeckel2007}.  We note that \citet{narayanan1996}
also suggested periodic outflow from Cep~A, although the dynamical timescale they determined
for the launch of successive CO shells is $\sim1.6\times10^5$ yr, two orders of magnitude greater than the
$\sim2500$ yr timescale suggested by the higher velocity shocked \hh{} gas presented here.

The second outflow from the Cep A core region appears to emerge from slightly
south of HW2.  This flow is associated with a low-velocity blueshifted lobe
of CO emission that terminates about 1.5 to 2\arcmin{} to the east in a faint but
large \hh{} bow shock.   HW3d is associated with strong \water{} masers
that are located at the base of a chain of radio sources at position angle
P.A. $\approx$ 105\arcdeg{}  \citep{torrelles1998, torrelles2001a} that has
the spectral index and morphology of a thermal radio jet that has the same
orientation as the suspected large-scale outflow from HW3c.

A third young star located southwest of HW2 may also power an outflow
that produced the radio sources HW1b, 3a, and 3b.    HW3b is the third source
in Cep A to contain masers \citep{torrelles1998, torrelles2001a}.    The 
submillimeter source SMA4 \citep{brogan2007} is located on the axis of
this chain close to HW3b.   The chain of radio sources has a position
angle P.A. $\approx$ 60--65\arcdeg{} (or 240 to 245\arcdeg ), close to the 
orientation of the \hh{} jet associated with the third suspected eruption 
of HW2.  Thus, an alternative to our model is that this collimated component
east of the Cep A core is a steady jet associated with a protostar located
in the HW1b, 3a, and 3b chain.   If SMA4 is the source, then the chain of radio 
sources is the counter-jet associated with the prominent \hh{} feature.
However, this model does not explain the orderly progression of terminal
bow-shock orientations and distances discussed above.   Furthermore, the
chain of radio sources HW1b, 3a, and 3b lies in the "line-of-fire" of the current
HW2 jet.  Thus, it is possible to interpret these features as shocks where the
HW2 jet rams part of the Cep A molecular cloud, or the outflow from HW3c.

No NIR continuum source is detected at the location of radio source W 
at the eastern end of HH~168.  The presence of radio continuum and X-ray
emission from this portion of HH~168 may indicate that this region is being
impacted by shocks with speeds of at least  400~\kms{}.   We propose 
that HH~168 marks the location where the HW3c outflow collides with the
outflow from HW2.    The location of HH~168 indicates that the currently shocked
gas is mostly excited by the debris from the third and fourth eruptions of HW2 
with the flow form HW3c.   Ejecta from the fifth eruption, which is traced by
the HW2 radio continuum jet, has not yet reached the interaction region.
If HW3b also drives its own jet with an orientation similar to the third
eruption from HW2, it may also contribute to the excitation of HH~168.

The complex shock morphology may be the result
of turbulent mixing of the colliding flows.  The north-to-south density gradient
in the ambient molecular cloud may contribute to the bending of
this shock complex toward the north.  The CO emission, which is both red- and
blueshifted at the location of HH~168, may also be deflected along our
line of sight by this gradient to produce the blueshifted CO lobe farther downstream.

Collisions between outflows and clouds have been invoked to explain
the complex structure of some Herbig--Haro objects such as HH 110 in
Orion where the HH~270 jet slams into a small molecular cloud in
the L1617 complex \citep{Lopez2005}.  An embedded IR source
in this cloud, IRAS 05487+0255,  drives a jet and  molecular outflow 
\citep{Reipurth_Olberg1991, Lee2000}.    Kajdic et al. (2009, in preparation) 
propose that HH~110 is formed by the collision of the HH~270 flow with the flow from
IRAS 05487+0255.    Theoretical models of jet--cloud interactions were
presented by \citet{CantoRaga1996, RagaCanto1996, deGouveiaDalPino1999,
Raga2002}.     Jet--jet collisions were modeled by \citet{Cunningham2006}.
In Cep A, HH~168 may be excited  by the collision between a 
pair of outflows and be additionally deflected by the density gradient in the
cloud.    Observations show that the sources of the HW2 and 3c flows are 
physically close.  Furthermore, both sources appear to drive flows that are 
close to the plane of the sky.  These factors make a collision of flows reasonably
probable.   The absence of obvious shocks southwest of the Cep A core supports
the view that ejecta from HW2 are deflected.     Still, the presence of low-velocity CO
emission southwest of HW2 and HW3c indicates that this part of the cloud has
been impacted by outflow activity.  A possible explanation for this dichotomy is
that,  in the distant past,  only the outflow from HW2 was active, and that the deflection
of this flow by the flow from HW3c is a relatively recent phenomenon.

\section{Conclusions}

We present new narrowband NIR images of the the Cep~A outflow
complex in the 2.12 $\mu$m S(1) line of \hh{}.   These images provide
evidence for two major outflows from the two most luminous members
of the Cep~A outflow complex.   

The radio source HW3c appears to drive a collimated east--west outflow 
with a redshifted lobe emerging toward P.A. $\approx$ 275\arcdeg{}  that
may be responsible for much of the emission associated with HH~168 in
Cep~A West.   The blueshifted lobe of this outflow appears to terminate in
a large but faint \hh{} bow shock located about 1.5--2\arcmin{}  east of
HW3c.   The \hh{} bow and the the brightest portion of HH~168 are placed
symmetrically about HW3c.  The three \hh{} bow shocks located along the
southern portion of HH~168 have orientations consistent with being driven by this source.

The radio source HW2, the most luminous and massive protostar in the
Cep~A complex, appears to drive a precessing and pulsed jet.  The \hh{} images 
reveal four distinct jet axes.  The longest and oldest outflow lobe terminates in 
HH~174 about 5\arcmin{} due east of HW2.   The second and third lobes are
progressively shorter, have rotated clockwise by about 10--15\arcdeg{} between
ejections, and terminate in the western and eastern  components of HH~169.
The fourth ejection is the shortest lobe and was ejected toward the northeast, 
but has no visual wavelength HH objects associated with it, probably due
to the high obscuration. The  radio continuum jet  may trace  a current period
of activity and continues the trend of clockwise rotation of the outflow axes.  
Thus, over the past 10$^4$ years, the HW2 outflow appears to have precessed
by nearly 45\arcdeg{}  in a clockwise direction as seen on the plane of the sky.
Assuming an average flow speed of 100~\kms{}, HW2 undergoes an eruption
about every 2500 years.

Most massive stars are born in clusters, and Cep~A is no exception.
In an environment with a high stellar density, interactions between a massive
star and its siblings become relatively common.  Cep~A HW2 is surrounded 
by a disk at least several hundred  AU in radius and with a mass of order 1 \msun{}.  
This disk  can serve as a  dissipative environment, and any star passing within 
about a  disk radius has a high probability of being captured into a highly
eccentric orbit.  The more massive the intruder, the higher its probability of 
capture \citep{moeckel2007a,moeckel2007}.  This mechanism provides a natural explanation 
for the presence of a moderate-mass companion, indicated in the HW2 system 
by a hot core 0.6\arcsec{} (400 AU in projection) east of HW2 \citep{martin-pintado2005}.

It is proposed that Cep~A HW2 has a moderate-mass companion in
an eccentric orbit whose orbital plane is inclined with respect to the
circumstellar disk surrounding HW2.   Periastron passages of the companion
may perturb the disk, and drive accretion onto the central star,  and produce
quasi-periodic episodes of collimated mass ejection.  The non-coplanar 
eccentric orbit also applies  a torque to the disk, changing its orientation.
We investigate this hypothesis using an SPH code and demonstrate that
such interactions can result in the type of disk orientation change proposed 
to have occurred in Cep~A\@.  Modeling indicates that a companion in a 
prograde orbit inclined with respect to the HW2 disk by about 40\arcdeg{}  
fits the observations best.   Because an impactor mass greater than the disk 
mass is needed to torque the disk through an appreciable angle, the 
presence of one or more low-mass sources in the vicinity do not affect this 
model. 

The bright shock complex HH~168  may
be excited by the collision of two or more outflows emerging from the
Cep A cloud core located about 90\arcsec{} to the east.  The western end
of HH~168 may trace the chaotic mixing resulting from the collision
of the steady (in orientation) flow from HW3c and the third eruption of HW2;
the eastern end of HH~168 may trace the collision of the flow from HW3c with
the fourth eruption of HW2.  If SMA4 also drives a flow, it may also
participate in this interaction.

\acknowledgments
This work was supported by NSF grant  AST0407356 and the CU Center
for  Astrobiology funded by NASA under cooperative 
agreement no. NNA04CC11A issued by the Office of Space Science.  Some of the observations presented
were obtained with the Apache Point Observatory 3.5-meter telescope,
which is owned and operated by the Astrophysical Research Consortium.  We thank
Drs.\ Mark Morris and Ralph Shuping for assistance with the acquisition and
reduction of the Keck data.

\bibliographystyle{apj}

\begin{thebibliography}{61}
\bibitem[{{Bally} {et~al.}(1995){Bally}, {Devine}, {Fesen}, \&
  {Lane}}]{bally1995}
{Bally}, J., {Devine}, D., {Fesen}, R.~A., \& {Lane}, A.~P. 1995, \apj, 454,
  345

\bibitem[{{Bally} \& {Lane}(1990)}]{bally1990}
{Bally}, J., \& {Lane}, A.~P. 1990, in ASP Conf. Ser. 14: Astrophysics with
  Infrared Arrays, ed. R.~{Elston} (San Francisco: ASP), 273

\bibitem[{{Bally} {et~al.}(2006){Bally}, {Licht}, {Smith}, \&
  {Walawender}}]{bally2006}
{Bally}, J., {Licht}, D., {Smith}, N., \& {Walawender}, J. 2006, \aj, 131, 473

\bibitem[{{Bally} \& {Reipurth}(2001)}]{bally01}
{Bally}, J., \& {Reipurth}, B. 2001, \apj, 546, 299

\bibitem[{{Blaauw} {et~al.}(1959){Blaauw}, {Hiltner}, \&
  {Johnson}}]{blaauw1959}
{Blaauw}, A., {Hiltner}, W.~A., \& {Johnson}, H.~L. 1959, \apj, 130, 69

\bibitem[{{Brogan} {et~al.}(2007){Brogan}, {Chandler}, {Hunter}, {Shirley}, \&
  {Sarma}}]{brogan2007}
{Brogan}, C.~L., {Chandler}, C.~J., {Hunter}, T.~R., {Shirley}, Y.~L., \&
  {Sarma}, A.~P. 2007, \apjl, 660, L133

\bibitem[{{Canto} \& {Raga}(1996)}]{CantoRaga1996}
{Canto}, J., \& {Raga}, A.~C. 1996, \mnras, 280, 559

\bibitem[{{Cohen} {et~al.}(1984){Cohen}, {Rowland}, \& {Blair}}]{cohen1984}
{Cohen}, R.~J., {Rowland}, P.~R., \& {Blair}, M.~M. 1984, \mnras, 210, 425

\bibitem[{{Colome} \& {Harvey}(1995)}]{colome1995}
{Colome}, C., \& {Harvey}, P.~M. 1995, \apj, 449, 656

\bibitem[{{Comito} {et~al.}(2007){Comito}, {Schilke}, {Endesfelder},
  {Jim{\'e}nez-Serra}, \& {Mart{\'{\i}}n-Pintado}}]{comito2007}
{Comito}, C., {Schilke}, P., {Endesfelder}, U., {Jim{\'e}nez-Serra}, I., \&
  {Mart{\'{\i}}n-Pintado}, J. 2007, \aap, 469, 207

\bibitem[{{Crawford} \& {Barnes}(1970)}]{crawford1970}
{Crawford}, D.~L., \& {Barnes}, J.~V. 1970, \aj, 75, 952

\bibitem[{{Cunningham} {et~al.}(2006){Cunningham}, {Frank}, \&
  {Blackman}}]{Cunningham2006}
{Cunningham}, A.~J., {Frank}, A., \& {Blackman}, E.~G. 2006, \apj, 646, 1059

\bibitem[{{Curiel} {et~al.}(2006){Curiel}, {Ho}, {Patel}, {Torrelles},
  {Rodr{\'{\i}}guez}, {Trinidad}, {Cant{\'o}}, {Hern{\'a}ndez}, {G{\'o}mez},
  {Garay}, \& {Anglada}}]{curiel2006}
{Curiel}, S., {Ho}, P.~T.~P., {Patel}, N.~A., {Torrelles}, J.~M.,
  {Rodr{\'{\i}}guez}, L.~F., {Trinidad}, M.~A., {Cant{\'o}}, J.,
  {Hern{\'a}ndez}, L., {G{\'o}mez}, J.~F., {Garay}, G., \& {Anglada}, G. 2006,
  \apj, 638, 878

\bibitem[{{Curiel} {et~al.}(2002){Curiel}, {Trinidad}, {Cant{\'o}},
  {Rodr{\'\i}guez}, {Torrelles}, {Ho}, {Patel}, {Greenhill}, {G{\'o}mez},
  {Garay}, {Hern{\'a}ndez}, {Contreras}, \& {Anglada}}]{curiel2002}
{Curiel}, S., {Trinidad}, M.~A., {Cant{\'o}}, J., {Rodr{\'\i}guez}, L.~F.,
  {Torrelles}, J.~M., {Ho}, P.~T.~P., {Patel}, N.~A., {Greenhill}, L.,
  {G{\'o}mez}, J.~F., {Garay}, G., {Hern{\'a}ndez}, L., {Contreras}, M.~E., \&
  {Anglada}, G. 2002, \apjl, 564, L35

\bibitem[{{Davis} {et~al.}(1997){Davis}, {Eisloeffel}, {Ray}, \&
  {Jenness}}]{davis97}
{Davis}, C.~J., {Eisloeffel}, J., {Ray}, T.~P., \& {Jenness}, T. 1997, \aap,
  324, 1013

\bibitem[{{de Gouveia Dal Pino}(1999)}]{deGouveiaDalPino1999}
{de Gouveia Dal Pino}, E.~M. 1999, \apj, 526, 862

\bibitem[{{Devine} {et~al.}(1997){Devine}, {Bally}, {Reipurth}, \&
  {Heathcote}}]{devine1997}
{Devine}, D., {Bally}, J., {Reipurth}, B., \& {Heathcote}, S. 1997, \aj, 114,
  2095

\bibitem[{{Eisloffel} {et~al.}(1996){Eisloffel}, {Smith}, {Davis}, \&
  {Ray}}]{eisloffel96}
{Eisloffel}, J., {Smith}, M.~D., {Davis}, C.~J., \& {Ray}, T.~P. 1996, \aj,
  112, 2086

\bibitem[{{Garay} {et~al.}(1996){Garay}, {Ramirez}, {Rodr{\'\i}guez}, {Curiel},
  \& {Torrelles}}]{garay1996}
{Garay}, G., {Ramirez}, S., {Rodr{\'\i}guez}, L.~F., {Curiel}, S., \&
  {Torrelles}, J.~M. 1996, \apj, 459, 193

\bibitem[{{Goetz} {et~al.}(1998){Goetz}, {Pipher}, {Forrest}, {Watson},
  {Raines}, {Woodward}, {Greenhouse}, {Smith}, {Hughes}, \&
  {Fischer}}]{goetz1998}
{Goetz}, J.~A., {Pipher}, J.~L., {Forrest}, W.~J., {Watson}, D.~M., {Raines},
  S.~N., {Woodward}, C.~E., {Greenhouse}, M.~A., {Smith}, H.~A., {Hughes},
  V.~A., \& {Fischer}, J. 1998, \apj, 504, 359

\bibitem[{{G{\'o}mez} {et~al.}(1999){G{\'o}mez}, {Sargent}, {Torrelles}, {Ho},
  {Rodr{\'\i}guez}, {Cant{\'o}}, \& {Garay}}]{gomez1999}
{G{\'o}mez}, J.~F., {Sargent}, A.~I., {Torrelles}, J.~M., {Ho}, P.~T.~P.,
  {Rodr{\'\i}guez}, L.~F., {Cant{\'o}}, J., \& {Garay}, G. 1999, \apj, 514, 287

\bibitem[{{Gutermuth} {et~al.}(2005){Gutermuth}, {Megeath}, {Pipher}, {Allen},
  {Williams}, {Allen}, {Myers}, \& {Fazio}}]{gutermuth2005}
{Gutermuth}, R.~A., {Megeath}, S.~T., {Pipher}, J.~L., {Allen}, T.~S.,
  {Williams}, J.~P., {Allen}, L.~E., {Myers}, P.~C., \& {Fazio}, G.~G. 2005, in
  Protostars and Planets V (Waikoloa, Hawaii), 8585

\bibitem[{{Hartigan} {et~al.}(1986){Hartigan}, {Lada}, {Tapia}, \&
  {Stocke}}]{hartigan1986}
{Hartigan}, P., {Lada}, C.~J., {Tapia}, S., \& {Stocke}, J. 1986, \aj, 92, 1155

\bibitem[{{Hartigan} {et~al.}(2000){Hartigan}, {Morse}, \&
  {Bally}}]{hartigan2000}
{Hartigan}, P., {Morse}, J., \& {Bally}, J. 2000, \aj, 120, 1436

\bibitem[{{Hiriart} {et~al.}(2004){Hiriart}, {Salas}, \&
  {Cruz-Gonz{\'a}lez}}]{hiriart2004}
{Hiriart}, D., {Salas}, L., \& {Cruz-Gonz{\'a}lez}, I. 2004, \aj, 128, 2917

\bibitem[{{Hoare} \& {Garrington}(1995)}]{hoare1995}
{Hoare}, M.~G., \& {Garrington}, S.~T. 1995, \apj, 449, 874

\bibitem[{{Hughes} {et~al.}(1995){Hughes}, {Cohen}, \&
  {Garrington}}]{hughes1995}
{Hughes}, V.~A., {Cohen}, R.~J., \& {Garrington}, S. 1995, \mnras, 272, 469

\bibitem[{{Hughes} \& {Wouterloot}(1984)}]{hughes1984}
{Hughes}, V.~A., \& {Wouterloot}, J.~G.~A. 1984, \apj, 276, 204

\bibitem[{{Jim{\'e}nez-Serra} {et~al.}(2007){Jim{\'e}nez-Serra},
  {Mart{\'{\i}}n-Pintado}, {Rodr{\'{\i}}guez-Franco}, {Chandler}, {Comito}, \&
  {Schilke}}]{jimenez-serra2007}
{Jim{\'e}nez-Serra}, I., {Mart{\'{\i}}n-Pintado}, J.,
  {Rodr{\'{\i}}guez-Franco}, A., {Chandler}, C., {Comito}, C., \& {Schilke}, P.
  2007, \apjl, 661, L187

\bibitem[{{Jones} \& {Puetter}(1993)}]{JonesPuetter1993}
{Jones}, B., \& {Puetter}, R.~C. 1993, \procspie, 1946, 610

\bibitem[{{Koppenaal} {et~al.}(1979){Koppenaal}, {van Duinen}, {Aalders},
  {Sargent}, \& {Nordh}}]{koppenaal1979}
{Koppenaal}, K., {van Duinen}, R.~J., {Aalders}, J.~W.~G., {Sargent}, A.~I., \&
  {Nordh}, L. 1979, \aap, 75, L1

\bibitem[{{Lee} {et~al.}(2000){Lee}, {Mundy}, {Reipurth}, {Ostriker}, \&
  {Stone}}]{Lee2000}
{Lee}, C.-F., {Mundy}, L.~G., {Reipurth}, B., {Ostriker}, E.~C., \& {Stone},
  J.~M. 2000, \apj, 542, 925

\bibitem[{{L{\'o}pez} {et~al.}(2005){L{\'o}pez}, {Estalella}, {Raga}, {Riera},
  {Reipurth}, \& {Heathcote}}]{Lopez2005}
{L{\'o}pez}, R., {Estalella}, R., {Raga}, A.~C., {Riera}, A., {Reipurth}, B.,
  \& {Heathcote}, S.~R. 2005, \aap, 432, 567

\bibitem[{{Mart{\'\i}n-Pintado} {et~al.}(2005){Mart{\'\i}n-Pintado},
  {Jim{\'e}nez-Serra}, {Rodr{\'\i}guez-Franco}, {Mart{\'\i}n}, \&
  {Thum}}]{martin-pintado2005}
{Mart{\'\i}n-Pintado}, J., {Jim{\'e}nez-Serra}, I., {Rodr{\'\i}guez-Franco},
  A., {Mart{\'\i}n}, S., \& {Thum}, C. 2005, \apjl, 628, L61

\bibitem[{{Moeckel} \& {Bally}(2006)}]{moeckel2006}
{Moeckel}, N., \& {Bally}, J. 2006, \apj, 653, 437

\bibitem[{{Moeckel} \& {Bally}(2007{\natexlab{a}})}]{moeckel2007a}
---. 2007{\natexlab{a}}, \apj, 656, 275

\bibitem[{{Moeckel} \& {Bally}(2007{\natexlab{b}})}]{moeckel2007}
---. 2007{\natexlab{b}}, \apj, 661, L183

\bibitem[{{Narayanan} \& {Walker}(1996)}]{narayanan1996}
{Narayanan}, G., \& {Walker}, C.~K. 1996, \apj, 466, 844

\bibitem[{{Papaloizou} \& {Terquem}(1995)}]{papaloizou95}
{Papaloizou}, J.~C.~B., \& {Terquem}, C. 1995, \mnras, 274, 987

\bibitem[{{Patel} {et~al.}(2005){Patel}, {Curiel}, {Sridharan}, {Zhang},
  {Hunter}, {Ho}, {Torrelles}, {Moran}, {G{\'o}mez}, \& {Anglada}}]{patel2005}
{Patel}, N.~A., {Curiel}, S., {Sridharan}, T.~K., {Zhang}, Q., {Hunter}, T.~R.,
  {Ho}, P.~T.~P., {Torrelles}, J.~M., {Moran}, J.~M., {G{\'o}mez}, J.~F., \&
  {Anglada}, G. 2005, \nat, 437, 109

\bibitem[{{Pravdo} \& {Tsuboi}(2005)}]{pravdo2005}
{Pravdo}, S.~H., \& {Tsuboi}, Y. 2005, \apj, 626, 272

\bibitem[{{Raga} \& {Canto}(1996)}]{RagaCanto1996}
{Raga}, A.~C., \& {Canto}, J. 1996, \mnras, 280, 567

\bibitem[{{Raga} {et~al.}(2002){Raga}, {de Gouveia Dal Pino}, {Noriega-Crespo},
  {Mininni}, \& {Vel{\'a}zquez}}]{Raga2002}
{Raga}, A.~C., {de Gouveia Dal Pino}, E.~M., {Noriega-Crespo}, A., {Mininni},
  P.~D., \& {Vel{\'a}zquez}, P.~F. 2002, \aap, 392, 267

\bibitem[{{Reipurth} \& {Bally}(2001)}]{reipurth2001}
{Reipurth}, B., \& {Bally}, J. 2001, \araa, 39, 403

\bibitem[{{Reipurth} \& {Olberg}(1991)}]{Reipurth_Olberg1991}
{Reipurth}, B., \& {Olberg}, M. 1991, \aap, 246, 535

\bibitem[{{Rodr{\'\i}guez} {et~al.}(1994){Rodr{\'\i}guez}, {Garay}, {Curiel},
  {Ramirez}, {Torrelles}, {Gomez}, \& {Velazquez}}]{rodriguez1994}
{Rodr{\'\i}guez}, L.~F., {Garay}, G., {Curiel}, S., {Ramirez}, S., {Torrelles},
  J.~M., {Gomez}, Y., \& {Velazquez}, A. 1994, \apjl, 430, L65

\bibitem[{{Rodr{\'\i}guez} {et~al.}(1980{\natexlab{a}}){Rodr{\'\i}guez}, {Ho},
  \& {Moran}}]{rodriguez1980a}
{Rodr{\'\i}guez}, L.~F., {Ho}, P.~T.~P., \& {Moran}, J.~M. 1980{\natexlab{a}},
  \apjl, 240, L149

\bibitem[{{Rodr{\'\i}guez} {et~al.}(1980{\natexlab{b}}){Rodr{\'\i}guez},
  {Moran}, {Ho}, \& {Gottlieb}}]{rodriguez1980}
{Rodr{\'\i}guez}, L.~F., {Moran}, J.~M., {Ho}, P.~T.~P., \& {Gottlieb}, E.~W.
  1980{\natexlab{b}}, \apj, 235, 845

\bibitem[{{Rodr{\'\i}guez} {et~al.}(2005){Rodr{\'\i}guez}, {Torrelles}, {Raga},
  {Cant{\'o}}, {Curiel}, \& {Garay}}]{rodriguez2005a}
{Rodr{\'\i}guez}, L.~F., {Torrelles}, J.~M., {Raga}, A.~C., {Cant{\'o}}, J.,
  {Curiel}, S., \& {Garay}, G. 2005, Rev. Mexicana Astron. AstroÞs., 41, 435

\bibitem[{{Sargent}(1977)}]{sargent1977}
{Sargent}, A.~I. 1977, \apj, 218, 736

\bibitem[{{Sargent}(1979)}]{sargent1979}
---. 1979, \apj, 233, 163

\bibitem[{{Shuping} {et~al.}(2004){Shuping}, {Morris}, \&
  {Bally}}]{shuping2004}
{Shuping}, R.~Y., {Morris}, M., \& {Bally}, J. 2004, \aj, 128, 363

\bibitem[{{Springel}(2005)}]{springel05}
{Springel}, V. 2005, \mnras, 364, 1105

\bibitem[{{Su} {et~al.}(2007){Su}, {Liu}, {Chen}, {Zhang}, \&
  {Cesaroni}}]{su2007}
{Su}, Y.-N., {Liu}, S.-Y., {Chen}, H.-R., {Zhang}, Q., \& {Cesaroni}, R. 2007,
  \apj, 671, 571

\bibitem[{{Terquem} {et~al.}(1999){Terquem}, {Eisl{\"o}ffel}, {Papaloizou}, \&
  {Nelson}}]{terquem99}
{Terquem}, C., {Eisl{\"o}ffel}, J., {Papaloizou}, J.~C.~B., \& {Nelson}, R.~P.
  1999, \apjl, 512, L131

\bibitem[{{Torrelles} {et~al.}(1998){Torrelles}, {G{\'o}mez}, {Garay},
  {Rodr{\'{\i}}guez}, {Curiel}, {Cohen}, \& {Ho}}]{torrelles1998}
{Torrelles}, J.~M., {G{\'o}mez}, J.~F., {Garay}, G., {Rodr{\'{\i}}guez}, L.~F.,
  {Curiel}, S., {Cohen}, R.~J., \& {Ho}, P.~T.~P. 1998, \apj, 509, 262

\bibitem[{{Torrelles} {et~al.}(2007){Torrelles}, {Patel}, {Curiel}, {Ho},
  {Garay}, \& {Rodr{\'{\i}}guez}}]{torrelles2007}
{Torrelles}, J.~M., {Patel}, N.~A., {Curiel}, S., {Ho}, P.~T.~P., {Garay}, G.,
  \& {Rodr{\'{\i}}guez}, L.~F. 2007, \apjl, 666, L37

\bibitem[{{Torrelles} {et~al.}(2001{\natexlab{a}}){Torrelles}, {Patel},
  {G{\'o}mez}, {Ho}, {Rodr{\'\i}guez}, {Anglada}, {Garay}, {Greenhill},
  {Curiel}, \& {Cant{\'o}}}]{torrelles2001a}
{Torrelles}, J.~M., {Patel}, N.~A., {G{\'o}mez}, J.~F., {Ho}, P.~T.~P.,
  {Rodr{\'\i}guez}, L.~F., {Anglada}, G., {Garay}, G., {Greenhill}, L.,
  {Curiel}, S., \& {Cant{\'o}}, J. 2001{\natexlab{a}}, \apj, 560, 853

\bibitem[{{Torrelles} {et~al.}(2001{\natexlab{b}}){Torrelles}, {Patel},
  {G{\'o}mez}, {Ho}, {Rodr{\'\i}guez}, {Anglada}, {Garay}, {Greenhill},
  {Curiel}, \& {Cant{\'o}}}]{torrelles2001b}
---. 2001{\natexlab{b}}, \nat, 411, 277

\bibitem[{{Torrelles} {et~al.}(1993){Torrelles}, {Verdes-Montenegro}, {Ho},
  {Rodr{\'\i}guez}, \& {Canto}}]{torrelles1993}
{Torrelles}, J.~M., {Verdes-Montenegro}, L., {Ho}, P.~T.~P., {Rodr{\'\i}guez},
  L.~F., \& {Canto}, J. 1993, \apj, 410, 202

\bibitem[{{Yu} {et~al.}(1999){Yu}, {Billawala}, \& {Bally}}]{yu1999}
{Yu}, K.~C., {Billawala}, Y., \& {Bally}, J. 1999, \aj, 118, 2940

\end{thebibliography}

\end{document}